# Quantifying the electrostatic transduction mechanism in carbon nanotube molecular sensors

Mitchell B. Lerner[1], James M. Resczenski[1], Akshay Amin[1], Robert R. Johnson[1], Jonas I. Goldsmith[2], A.T. Charlie Johnson[1]

[1]Department of Physics and Astronomy, University of Pennsylvania, 209 South 33rd Street, Philadelphia, PA 19104
[2]Department of Chemistry, Bryn Mawr College, 101 North Merion Avenue, Bryn Mawr, Pennsylvania 19010

Supporting Information Placeholder

**ABSTRACT:** Despite the great promise of carbon nanotube field effect transistors (CNT FETs) for applications in chemical and biochemical detection, a quantitative understanding of sensor responses is lacking. To explore the role of electrostatics in sensor transduction, experiments were conducted with a set of highly similar compounds designed to adsorb onto the CNT FET *via* a pyrene linker group and take on a set of known charge states under ambient conditions. Acidic and basic species were observed to induce threshold voltage shifts of opposite sign, consistent with gating of the CNT FET by local charges due to protonation or deprotonation of pyrene compounds by interfacial water. The magnitude of the gate voltage shift was controlled by the distance between the charged group and the CNT. Additionally, functionalization with an uncharged pyrene compound showed a threshold shift ascribed to its molecular dipole moment. This work illustrates a method to produce CNT FETs with controlled values of the turnoff gate voltage, and more generally, these results will inform the development of quantitative models for the response of CNT FET chemical and biochemical sensors.

Single walled carbon nanotubes have shown great promise for use as chemical sensors. Various surface modifications have been used to create nano-enabled, all-electronic vapor sensors,[1,2] electrochemical cells for small molecule detection,[3,4] and fast electronics-based protein detection.[5-7] However, the detection mechanisms for these devices remain incompletely understood. Pyrene-containing compounds have been shown to absorb specifically onto carbon nanotubes through π-π stacking, and this absorption process has been measured *in situ* by electrochemical methods.[8] This makes pyrene compounds an ideal system for exploring transduction mechanisms since they provide a method to precisely position known chemical groups with respect to the nanotube sidewall.

Here, single walled carbon nanotube field effect transistors (CNT FETs) were functionalized the pyrene compounds shown in Fig. 1. The turnoff threshold voltage (*i.e.*, the backgate voltage required to suppress conduction in the FET) for the functionalized devices was measured and was found to shift as the acid-base properties of the pyrene molecules were varied. This shift was attributed to chemical (electrostatic) gating[9] of the CNT FET by protonated/deprotonated groups on the pyrene molecules. The size of the threshold voltage shift was controlled by the distance between the charged group and the CNT sidewall, leading to insights about the nature of the electrostatic interaction. Interestingly, a neutral compound induced a threshold gate voltage shift due its intrinsic dipole moment. The magnitude of the observed threshold voltage shifts were in agreement with quantitative estimates of the charge density associated with the adsorbed pyrene molecules. The work illustrates a practical functionalization scheme and shows how one could tailor the threshold voltage of CNT FETs by judicious choice of a compound. These results also help to build a quantitative understanding of electrostatic detection mechanisms in CNT FET molecular sensors.

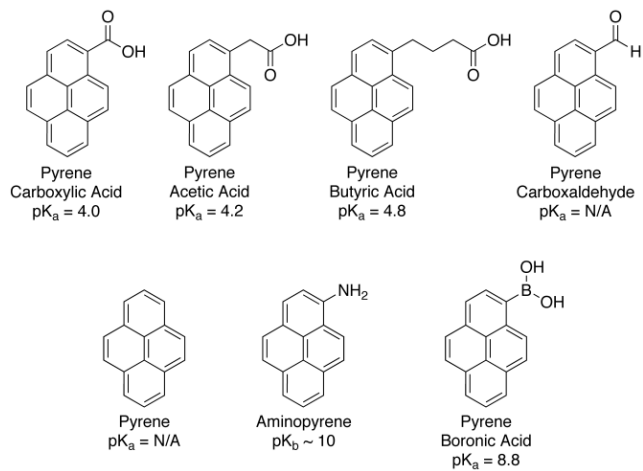

**Figure 1.** Pyrene compounds utilized in this work. Quoted values of $pK_a$ and $pK_b$ are from references [10-14]. Where actual values of $pK_a$ were not avaliable, values were estimated by comparison to related compounds: naphthylacetic acid for pyrene acetic acid, naphthylamine for aminopyrene, and phenylboronic acid for pyrene boronic acid.

Pyrene, 1-pyrene carboxylic acid, 1-aminopyrene and 1-pyrene acetic acid were purchased from Sigma Aldrich; 1-pyrene butyric acid and 1-pyrene carboxaldehyde were purchased from Alfa Aesar; and 1-pyrene boronic acid was purchased from TCI. The solvent used for all solutions was Honeywell Burdick & Jackson Brand high purity (99.9% pesticide residue grade) acetonitrile ($CH_3CN$). All chemicals were used as received. To make solutions of various pyrenes, a concentrated stock solution (*ca*. 1 mM) was prepared by dissolving a small amount (*ca*. 5 mg) of each pyrene in 20.0 mL of $CH_3CN$. Several minutes of mixing using an ultrasonic bath were required to effect complete dissolution. Stock solutions were diluted to yield 5 µM pyrene solutions in $CH_3CN$.

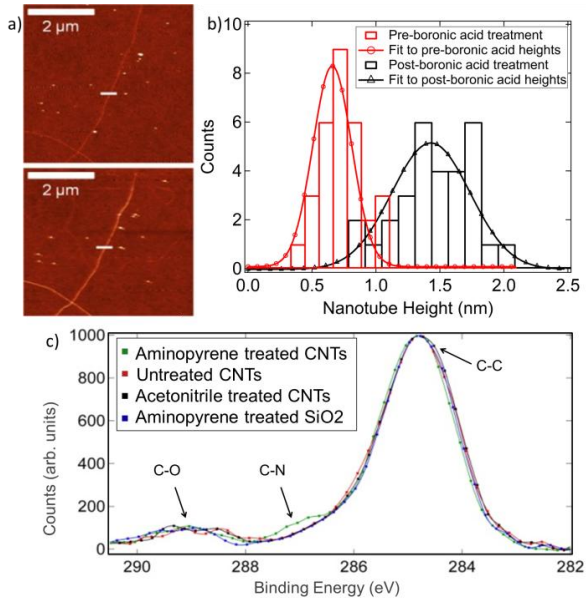

**Figure 2.** a) AFM images of a CNT before (top) and after (bottom) treatment with 5 μM pyrene boronic acid. Z scale is 6 nm. b) Histogram based on 30 line scans at the same positions before and after pyrene boronic acid treatment. An increase in height of *ca.* 0.6 nm is evident, and attributed to the presence of adsorbed pyrene molecules. c) XPS spectra from several experimental conditions. The peak at 287.0 eV indicates the presence of C-N bonds in the sample of CNTs treated with aminopyrene.

Carbon nanotube FETs were fabricated as described previously.[15] Briefly, CNTs were grown on Si/SiO$_2$ substrates by chemical vapor deposition (CVD) at 900°C with iron nanoparticles as the catalyst. Source-drain electrodes were patterned by photolithography and metalized with 3 nm Ti/40 nm Pd deposited in a thermal evaporator. After liftoff, devices were annealed in ambient at 250°C for one hour to remove excess photoresist residue.[15] The underlying doped silicon substrate served as a global backgate in a three-terminal FET geometry.

CNT FETs were electrically characterized by measuring source-drain current as a function of gate voltage (I-V$_G$) at a bias voltage of 100 mV. Devices were found to be dominated by hole conduction (see Figure 3); only devices with ON/OFF ratios exceeding 100 were used in subsequent experiments. CNT FETs were functionalized by incubation in a solution of one of the pyrene compounds (5 μM in CH$_3$CN) for two hours, rinsed for five minutes each in CH$_3$CN, isopropanol, and deionized water, and then blown dry with clean nitrogen. Samples were baked on a hot plate at 120°C for two hours to remove any remaining solvents. Between 15 and 20 devices were functionalized by this process with each pyrene compound shown in Fig. 1.

Effectiveness of the functionalization procedure was confirmed *via* Atomic Force Microscopy (AFM) and X-ray Photoelectron Spectroscopy (XPS). AFM images (Fig. 2a,b) showed that the height of the CNTs increased by *ca.* 0.5 nm after functionalization, consistent with what would be expected from the interlayer spacing of graphitic systems[16] such as a CNT and an absorbed pyrene compound. The XPS spectrum of a device functionalized with aminopyrene (Fig. 2c) showed a peak at 287.0 eV, indicative of the C-N bond of the aminopyrene bound on the CNT, as well as peaks at 284.8 eV and 289.0 eV, characteristic of C-C and C-O bonding, respectively. Consistent with expectations, samples of untreated CNTs, acetonitrile treated CNTs, and bare SiO$_2$ treated with aminopyrene did not show the C-N peak (Fig. 2c).

Electrical measurements were performed after functionalization, and changes in the I-V$_G$ curve were noted, especially changes in the transistor threshold voltage (ΔV$_T$). Functionalization with pyrene compounds had little effect on CNT FET ON state resistance and hole carrier mobility, consistent with the intuition that non-covalent functionalization does not introduce significant carrier scattering (see Fig. 3). As discussed below, the results can be explained quantitatively in a picture where the value of ΔV$_T$ is determined by electrostatic chemical gating of the CNT FET by the adsorbed pyrene molecules.

Since a thin surface water layer exists on CNT FET devices on SiO$_2$ under ambient, its effect on the charge state of the adsorbed pyrene molecules must be considered. Recent work has demonstrated that silanol groups on the surface of SiO$_2$ cause near-surface water to be considerably more acidic than bulk water, by *ca.* 2 pH units[17]. Pyrene-functionalized CNT FET devices on SiO$_2$ are thus best understood as operating in a slightly acidic aqueous medium (pH~5).

The observed values of ΔV$_T$ are summarized in Fig. 4; they were reproducible for each of the pyrene compounds that were tested. Adsorption of pyrene on the CNT FET led to a statistically insignificant shift in V$_T$ (-0.2 ± 0.4 V); a similarly small shift was observed for pyrene carboxaldehyde (0.4 ± 0.4 V). These findings are consistent with the expectation that both compounds are in a neutral charge state under the experimental conditions, with little or no effect on the I(V$_G$) characteristic of the FET. The sign of ΔV$_T$ was positive for each of the three pyrene carboxylic acids tested and negative for aminopyrene. These observations are consistent with the equilibrium constants listed in Fig. 1. As the pK$_a$ values for all of the pyrene carboxylic acids are less than 5, most of the carboxylic acid functional groups should be deprotonated and negatively charged. The I(V$_G$) characteristic is thus expected to shift to more positive gate voltage ($\Delta V_T > 0$). Similarly, for aminopyrene at pH 5, a fraction (*ca.* 10-15%) of the amine groups will be protonated, leading to a negative value of ΔV$_T$. The case of pyrene boronic acid is unusual and is discussed later.

Interpretation of the data is informed by a calculation of *N*, the density of singly charged species adsorbed to the CNT sidewall (presumed to be pyrene molecules) required to produce an observed value of ΔV$_T$: $N = |C \Delta V_T / e|$, where *C* is the CNT capacitance per unit length, and *e* is the electron charge. The capacitance per unit length of a CNT FET in the backgate geometry is: $C = 2\pi\varepsilon\varepsilon_0 / \ln(2h/r)$, where *r* is the CNT radius, and *h* and ϵ are the thickness and dielectric constant of the insulator between the CNT and the backgate, respectively.[18] For the device geometry used here, this equates to approximately 200 charged molecules/μm per volt of ΔV$_T$.

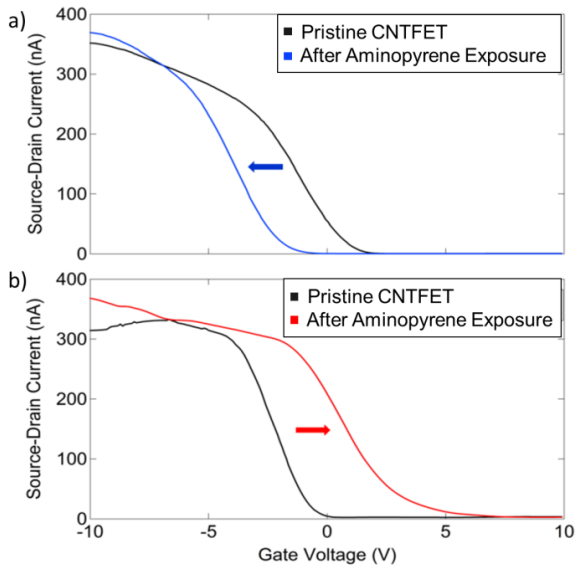

**Figure 3.** (a) I-$V_g$ plot for a CNT FET functionalized with aminopyrene shows a shift to negative $V_G$. (b) Similar data for a device functionalized with pyrenecarboxylic acid shows a shift to positive $V_G$.

The observed threshold voltage shift for pyrene carboxylic acid, $\Delta V_T = +5.0 \pm 0.7 V$, would require 1000 negative charges/μm of CNT length. By geometrical stacking arguments, full monolayer coverage of pyrenes comprises *ca.* 9000 molecules/μm. A lower bound of 3000 molecules/μm is suggested by recent measurements of the coverage for pyrenes coupled to transition metal complexes that are significantly larger than the functional groups of the molecules used here.[8] For this discussion, a density of 6000/μm is assumed. In a water layer with pH 5, roughly 90% (*i.e.* ~5000/μm) of the carboxylic acid functional groups will be deprotonated. Since the calculation suggests that 1000 deprotonated groups/μm are needed to produce the observed threshold voltage shift, we conclude that the gating effect of the charged groups is reduced by a factor of about 5 due to screening. For aminopyrene ($\Delta V_T = -2.8 \pm 0.3 V$), a similar density of adsorbed molecules is expected, but only 15% of the molecules should be charged, *i.e.* 4.5 times fewer than for pyrene carboxylic acid. This argument leads to a predicted threshold voltage shift of -1.1 V, about a factor of 2.5 less than what is observed. This could be explained by a difference in the screening effect in the two cases or an environment with pH lower than 5, which would result in a smaller fraction of deprotonated carboxylic acid groups and larger fraction of protonated amino groups. With these caveats, this framework provides a satisfactory semi-quantitative explanation of the observations.

It was also observed that varying the compound containing a carboxyl group (carboxylic acid to acetic acid to butyric acid) led to a systematic reduction in the measured value of $\Delta V_T$ ($+5.0 \pm 0.7 V$, $+1.2 \pm 0.3 V$, $+0.1 \pm 0.4 V$, respectively). This progression is attributed primarily to successively larger displacement of the charged carboxyl group away from the CNT sidewall; a secondary contributing factor is the reduced probability for deprotonation of the carboxyl group due to differing values of pKa for the compounds. Molecular dynamics (MD) simulations (see Supporting Information for details) indicated that the distance between the CNT sidewall and the carboxyl group of pyrene carboxylic acid is 0.34 nm, approximately equal to the interlayer spacing for graphene. This is interpreted as the effective electrostatic radius of the CNT and the pyrene molecule each extending 0.17 nm into the adjoining space, so the relevant distance for electrostatic influence (distance between centers of electron charge distributions) is 0.17 nm. The carboxyl group in pyrene acetic acid is 0.12 nm farther from the CNT than for pyrene carboxylic acid. Assuming Coulomb interactions between the carboxyl group and the CNT, and accounting for reduced protonation caused by the change of 0.2 units in pKa, this increased distance corresponds to a reduction in the interaction strength by a factor of 3.2, as compared to the observed reduction in $\Delta V_T$ by a factor of $4.1 \pm 1.3$. MD simulations indicated that the carboxyl group on pyrene butyric acid is 0.18 nm farther away than for pyrene carboxylic acid. This distance change coupled to the change in pKa corresponds to a reduction by a factor of 4.3 and a predicted threshold shift of roughly 0.7 V. The measured shift for butyric acid is smaller ($+0.1 \pm 0.4 V$), possibly due to more effective screening by water molecules that more readily penetrate between the carboxylic acid group and the nanotube surface for this molecule.

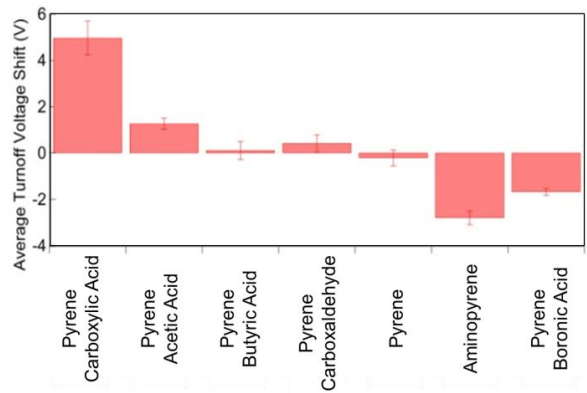

**Figure 4.** Average threshold voltage shift, $\Delta V_T$, caused by exposure to pyrene compounds. Each value is the average of results from 15-20 devices; error bars are the standard errors of the mean.

In contrast to other acids tested, functionalization with pyrene boronic acid led to a negative value of $\Delta V_T = -1.7 \pm 0.1 V$; that is, the *sign* of the threshold voltage shift was characteristic of a base rather than an acid. To explain this, two facts are noted: 1) pyrene boronic acid is a weak acid (p$K_a$ = 8.8), so at pH 5 virtually all of the boronic acid molecules are expected to be charge neutral; and 2) the B-O bonds in boronic acid are highly polar, with the boron (oxygen) atoms bearing a partial positive (negative) charge; thus, pyrene boronic acid acts as an adsorbed dipole. Quantum chemical simulations indicated a molecular dipole moment *ca.* 2.6 D, nearly coplanar with pyrene group. Assuming random orientation of the pyrene when adsorbed onto the CNT, on average the dipole will be oriented with the positively charged boron atom located closer to the CNT sidewall than the negatively charged oxygen by a distance of approximately 0.05Å. Based on these considerations, it is expected that the threshold voltage shift would be *negative* and approximately one-third that observed for the aminopyrene case ($\Delta V_T = -2.8 \pm 0.3 V$). The observed value of the threshold voltage shift, $\Delta V_T = -1.7 \pm 0.1 V$, is thus attributed to the electrostatic ef-fect of this dipole. The near-zero values of $\Delta V_T$ observed for pyrene and pyrene carboxaldehyde are consistent with this explanation since neither of those species contains a bond as polarized as the B-O bonds in pyrene boronic acid.

To summarize, we demonstrated that pyrene compounds, which specifically absorb onto CNT FETs via a π-π stacking interaction, can modify the transistor $I(V_G)$ characteristic through an electrostatic gating process. Acidic or basic functional groups interact with the interfacial water present on the CNT FET and become charged. These charged species, anchored onto the CNT, gate the device, modifying the nanotube surface potential through the electrostatic interaction. The change in surface potential results in a positive (negative) shift of the turnoff threshold voltage for negatively (positively) charged functional groups. Water-mediated local gating is thus a primary mechanism behind the responses of CNT FET molecular sensors. The magnitude of the threshold voltage shift was found to depend sensitively on the distance between the charged group and the CNT sidewall, behavior that can be well understood using an electrostatic model. The behavior of CNT FETs functionalized with uncharged pyrene boronic acid is attributed to the effect of the molecular dipole moment, suggesting an explanation for reports in the literature where exposure to neutral molecules was found to elicit a distinct response from CNT FET sensors.[19,20]

Future work will involve experiments in buffer solution in order to modify the local pH at the $SiO_2$/water interface. This should modulate the degree of protonation for a given species and lead to threshold voltage shifts. The data presented here and obtained through the further investigations will inform the development of MD-based simulations that go beyond existing structural studies[21-23] to calculate the response of CNT FET chemical sensors due to electrostatic interactions.

## ASSOCIATED CONTENT

**Supporting Information**. CNT FET schematic. Information on Molecular Dynamics computer simulations. This material is available free of charge via the Internet at http://pubs.acs.org

## AUTHOR INFORMATION

**Corresponding Author**

cjohnson@physics.upenn.edu, jigoldsmit@brynmawr.edu

**Notes**

The authors declare no competing financial interest.

## ACKNOWLEDGMENT

This research was supported by the by the Department of Defense US Army Medical Research and Materiel Command through grant W81XWH-09-1-0206 and by the Nano/Bio Interface Center through the National Science Foundation NSEC DMR08-32802. J.I.G. acknowledges support by a Faculty Start-up Award by the Camille and Henry Dreyfus Foundation. This work made use of the XPS/Auger system at the Centralized Research Facilities at Drexel University. The acquisition of the XPS/Auger system was made possible through the support of the National Science Foundation under Award CBET-0959361.

Table of Contents Graphic

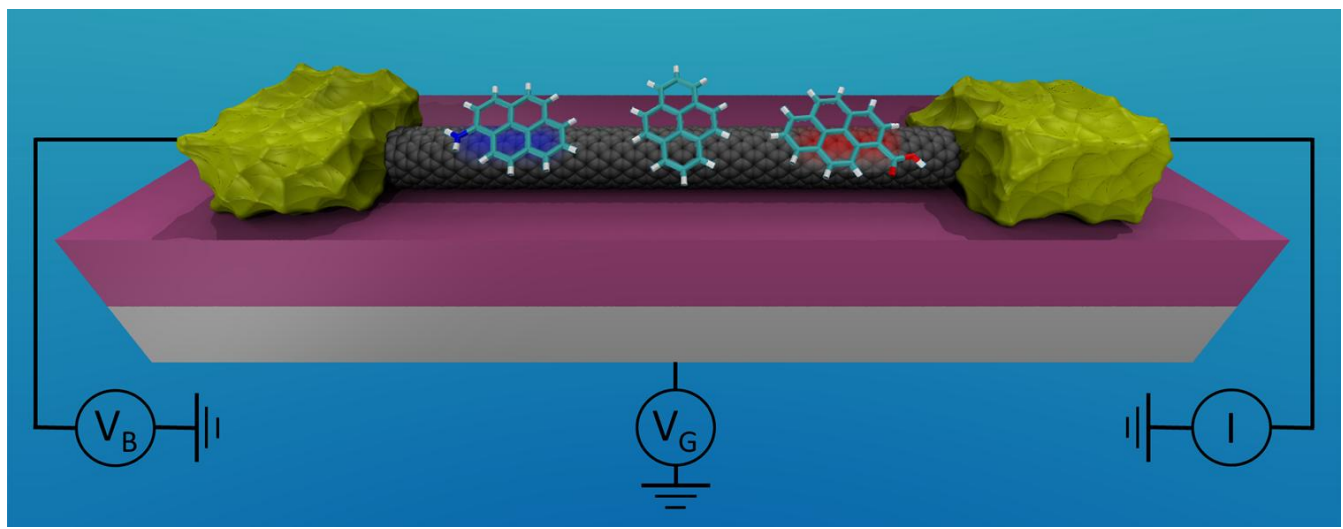